 \documentclass[final,authoryear,5p,times,twocolumn]{elsarticle}

\usepackage{graphics}
\usepackage{graphicx}
\usepackage{epsfig}
\usepackage[nodots,nocompress]{numcompress}

\usepackage{amssymb}
\usepackage{enumerate}
\usepackage{amsthm}
\usepackage{float}
\usepackage{lineno}
\usepackage{fancyhdr}

\journal{J. Atmos. Solar-Terr. Phys., 2014 (DOI: http://dx.doi.org/10.1016/j.jastp.2014.11.005)}

\begin{document}

\begin{frontmatter}



\title{Influence of cosmic-ray variability on the monsoon rainfall and temperature}


\author{Badruddin \& Aslam, O.P.M.}

\address{Aligarh Muslim University, Aligarh, India-202 002}

\begin{abstract}
We study the role of galactic cosmic ray (GCR) variability in influencing the rainfall variability  in Indian Summer Monsoon Rainfall (ISMR) season. We find that on an average during 'drought' (low ISMR) periods in India, GCR flux is decreasing, and during 'flood' (high ISMR) periods, GCR flux is increasing. The results of our analysis suggest for a possibility that the decreasing GCR flux during the summer monsoon season in India may suppress the rainfall. On the other hand, increasing GCR flux may enhance the rainfall. We suspect that in addition to real environmental conditions, significant levitation/dispersion of low clouds and hence reduced possibility of collision/coalescence to form raindrops suppresses the rainfall during decreasing GCR flux in monsoon season. On the other hand, enhanced collision/coalescence efficiency during increasing GCR flux due to electrical effects may contribute to enhancing the rainfall. Based on the observations, we put forward the idea that, under suitable environmental conditions, changing GCR flux may influence precipitation by suppressing/enhancing it, depending upon the decreasing/increasing nature of GCR flux variability during monsoon season in India, at least. We further note that the rainfall variability is inversely related to the temperature variation during ISMR season. We suggest an explanation, although speculative, how a decreasing/increasing GCR flux can influence the rainfall and the temperature. We speculate that the proposed hypothesis, based on the Indian climate data can be extended to whole tropical and sub-tropical belt, and that it may contribute to global temperature in a significant way. If correct, our hypothesis has important implication for the sun - climate link. 
\end{abstract}


\end{frontmatter}



\section{Introduction}
\label{Introduction}
The Asian summer monsoon is the largest single abnormality in the global climate system (Shukla, 2007). The seasonal rainfall brought by the southwest Indian summer monsoon supplies 80\% of Southeast Asia's annual precipitation and is vital to sustaining the region's agriculture which supports nearly a quarter of the world's population (Sinha et al., 2007). Indian summer monsoon is one of the main weather systems on earth and variations in its intensity have broad economic effects. It has been the most important climate event in India. Rainfall over India is subject to a high degree of variations leading to the occurrence of extreme monsoon rainfall deficient (drought) or excess (flood) over extensive areas of the country. Floods and droughts result in many losses of lives, crops etc.; these play havoc to Indian economy and society. 

Cause of abnormal variabilities in monsoon rainfall (floods and droughts) is not completely understood. Consequently, accurate prediction of rainfall and its variability during monsoon season has been a challenging task. Thus, there is greater need to understand the nature and variability of monsoon climatic conditions, especially, whether there is any extra-terrestrial influence (e.g. cosmic ray variability) in addition to natural terrestrial climatic conditions. More specifically, it is important to know whether Indian monsoon rainfall is significantly influenced by changes in cosmic ray flux, and whether climate cooling is an effect of cosmic ray flux change. If so, then the possible physical mechanism(s) must be identified.

It is well known that cosmic ray flux varies in anti-phase with solar activity over all time scales. On the longer time scales (millennial, centennial and multi-decadal), a number of studies have suggested solar/cosmic ray variability influence on the intensity of monsoonal rainfall in tropical and sub-tropical regions with conflicting results. For example, low rainfall in India coinciding with low solar activity (or high cosmic ray intensity) (e.g. Agnihotri et al., 2002; Tiwari et al., 2005; Gupta et al., 2005; Yadava and Ramesh, 2007) and in North Africa and South Oman (Neff et al., 2001). These results imply that increased galactic cosmic ray (GCR) intensity is associated with a weakening of the monsoon (decreased rainfall) (Kirkby, 2007; Singh et al., 2011). In contrast, low rainfall in equatorial East African (e.g. Verschuren et al., 2000), weaker Chinese monsoon (Hong et al., 2001), and low tropical rainfall in Gulf of Mexico region have been observed, during high solar activity (or low cosmic ray intensity). Occurrence of periods of enhanced monsoonal precipitation in India slightly after the termination of the Wolf, Sporer and Maunder minima periods (low solar activity/high cosmic ray intensity) have been reported by Khare and Nigam (2006). This finding is in agreement with the finding of earlier workers, who reported high lake levels from Mono Lake and Chad Lake in the vicinity of solar minima (cosmic ray maxima) as well as the Nile River in Africa (Ruzmaikin et al., 2006). Thus there are evidences, although sometimes contrary in nature, that suggest for some cosmic ray influence on monsoon rainfall on multi-decadal, centennial and millennial time scale.

On shorter time scales (decadal to inter annual) too, solar activity/cosmic ray intensity influence on the rainfall changes in Indian summer monsoon have been suggested, but with conflicting results (e.g. see Jagannathan and Bhalme, 1973; Bhalme et al., 1981; Hiremath and Mandi, 2004; Bhattacharya and Narasimha, 2005; Badruddin et al., 2006, 2009).

Understanding the factors that control ISMR onset, its variability and intensity are highly desired. In particular, it is extremely important to know about the role of extra-terrestrial sources (e.g. cosmic rays) in initiating and/or influencing the intensity of rainfall directly (e.g. by changing the collision/coalescence efficiency in rain clouds) or indirectly (e.g. by altering the low cloud amount). It is particularly important to search for connection, if any, between the extreme deficiency (droughts) or excess (floods) in Indian summer monsoon rainfall and cosmic ray flux variability during the same ISMR periods, even though it is widely accepted that Indian monsoon onset and intensity are controlled by large scale atmospheric (e.g. land-sea temperature contrast) and global features (e.g. ENSO, QBO etc.). 

Several studies have shown that the warm phase (El Nino) is associated with weakening of Indian monsoon with overall reduction in rainfall while the cold phase (La Nina) is associated with the strengthening of the Indian monsoon with enhancement in rainfall (e. g., Sikka, 1980; Pant and Parthasarathy, 1981; Rasmusson and Carpenter, 1983). All the El Nino events during 1958-1988 were reported to be droughts and all the La Nina events were associated with excess ISMR. However, weakening of ENSO-ISMR relationship after 1988 were reported in later studies (e. g., Kripalani and Kripalani, 1997; Kumar et al., 1999; Ashok et al., 2001; Kripalani et al., 2003). Further, for the 14 consecutive years beginning with 1988 (1988 to 2001), there were no droughts, despite the occurrence of El Nino (Gadgil et al., 2004). Although 9 out of 12 drought years identified by us can be associated with El Nino events, and 9 out of 12 flood years with La Nina events, there are reports (Kumar et al., 2002) that out of 22 large negative ISMR anomalies that occurred during 1871-2001, only 11 were associated with El Nino, while out of 19 large positive ISMR anomalies that occurred during the same period, only 8 were associated with La Nina. Therefore, large deficient/excess ISMR does occur in the absence of El Nino/La Nina and we do not yet understand adequately the response of monsoon to El Nino (Gadgil et al., 2004). Thus, there is the possibility of drought/floods in India being influenced by other external agents also.

\begin{figure}
\centerline{\includegraphics[width=\hsize]{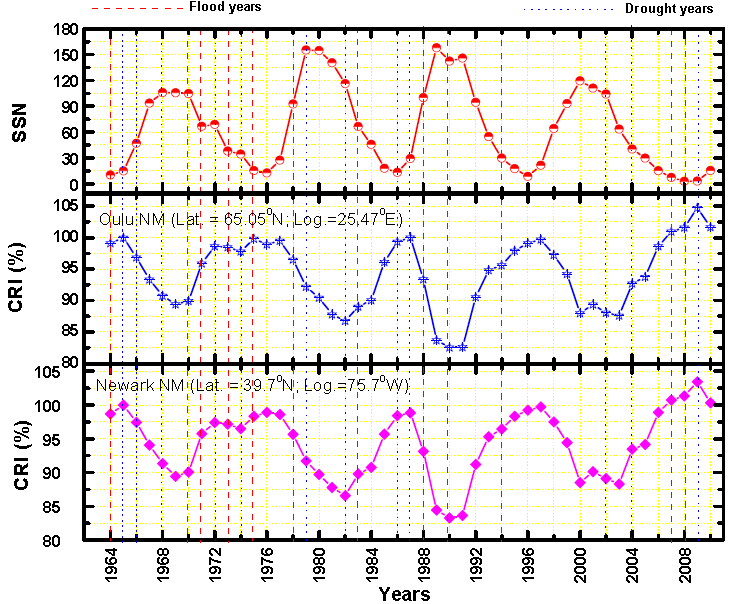}}
\caption{Yearly averaged sunspot number (SSN) and galactic cosmic ray (GCR) intensity variations as observed at Oulu and Newark neutron monitors. Dashed and dotted vertical lines representing drought and flood years respectively}
\end{figure}

Cosmic rays are the only source of ion production in the lower atmosphere. There are suggestions that cosmic ray flux variability may influence the earth's climate also. In view of these suggestions, although controversial, it will be interesting to search for any possibility of a link between GCR and rainfall variability. Variations in precipitation potentially caused by changes in the cosmic ray flux have implications for the understating of the cloud and water vapour feedbacks. It is possible that any particular (e.g. Indian) climate system is more sensitive to smaller variations in cosmic ray intensity than the other.

The purpose of this investigation is to determine the relationship, if any, between the Indian extreme weather (Drought/Flood) and cosmic ray flux variability. We analysed the GCR flux data to evaluate the possible existence of empirical evidence between cosmic ray variability and precipitation in India during monsoon season.

For this study we utilized the GCR fluxes as recorded through the ground based neutron monitors, and perform analysis to look for any possibility of changes in pattern in Indian rainfall, in particular, due to variations in GCR flux. For this purpose we adopt the methods of superposed epoch analysis (Singh and Badruddin, 2006) and regression analysis. We find evidence for a possibility that GCR flux variability may have some influence in suppressing/enhancing the rainfall depending upon the decreasing/increasing nature of GCR variability, in favourable climatic conditions.

\begin{figure*}[htp!]
   \centerline{\includegraphics[width=\hsize]{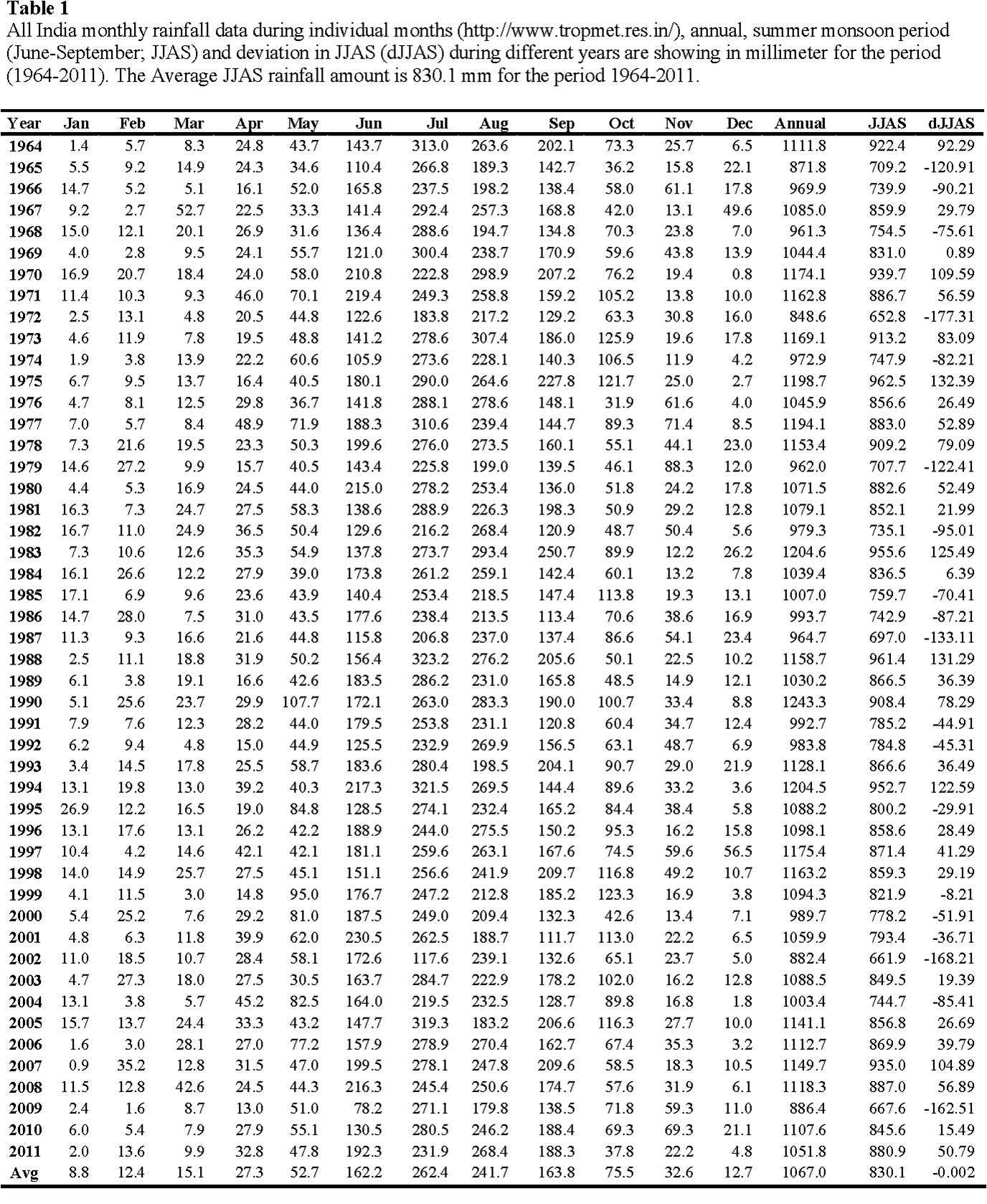}}
        \end{figure*}   

\section{Analysis}
\label{Analysis}

In this work we adopt an approach that assumes that the rainfall changes can occur only with GCR changes if environmental conditions are suitable, and considering that the rate of GCR flux change, and not the mean GCR flux, may be the key (Laken et al., 2010). Usoskin (2011), in a recent review, concluded that it is not the intensity of cosmic rays but its variability that may affect climate.

\begin{figure}[htp!]
   \centerline{\includegraphics[width=\hsize]{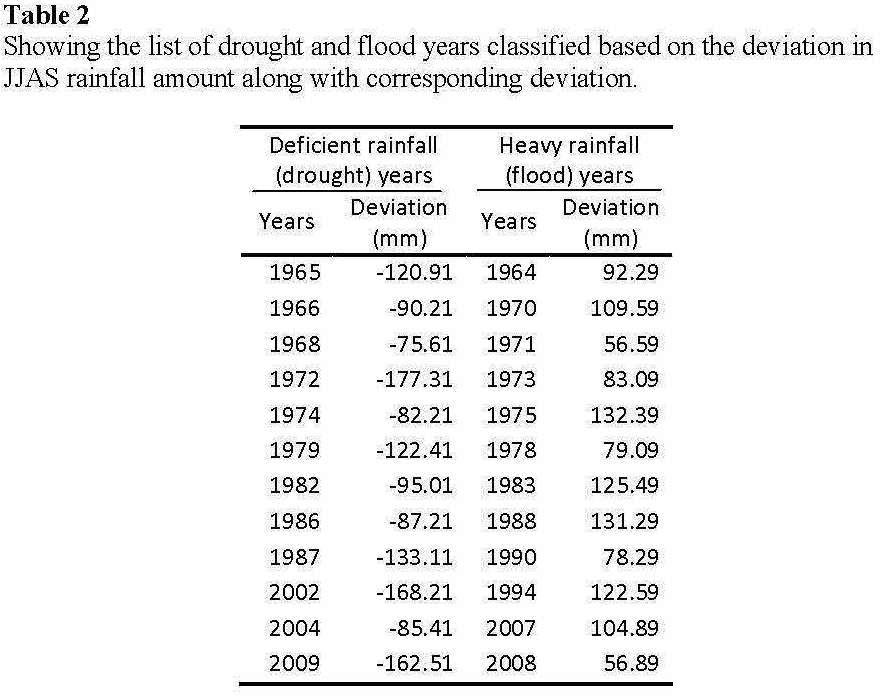}}
        \end{figure} 

The GCR flux is provided by neutron monitors, which record neutrons generated chiefly by the primary cosmic ray protons that ionize the lower stratosphere and upper troposphere (Venkatesan and Badruddin, 1990; Bazilevskaya and Svirzhevskaya, 1998). Continuous records of high quality cosmic ray intensity data, measured by neutron monitors located at different latitudes and longitudes on the earth's surface are available from 1964 onwards till date. Reliable and good quality data of monsoon rainfall in India are also available for the period 1964 - 2011 (see Table 1) and many more years before that, at Indian Institute of Tropical Metrology, Pune (India) website (http://www.tropmet.res.in/). For this work, we have considered the 48-year period (1964 - 2011) for which both the GCR intensity and Indian Summer Monsoon Rainfall (ISMR) data are available. 

\begin{figure}
\centerline{\includegraphics[width=\hsize]{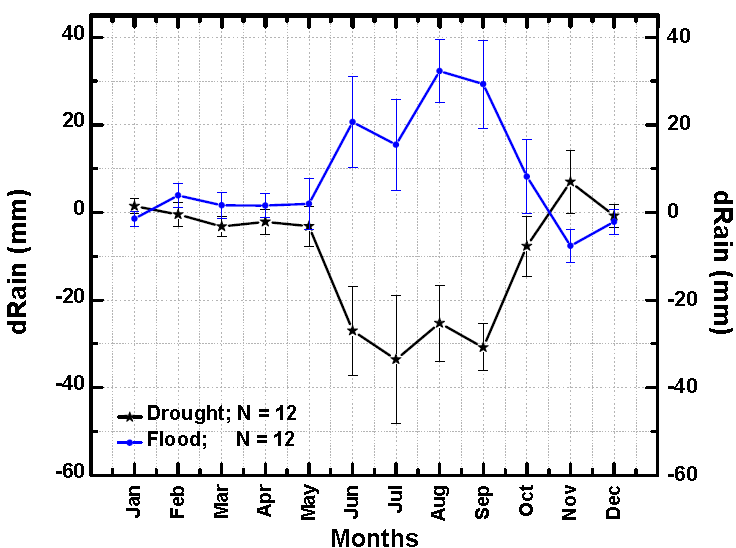}}
\caption{Deviation of monthly rainfall during superposed drought and flood years from superposed average rainfall of 48 years (1964-2011).}
\end{figure} 

\section{Results}
\label{Results}
A variation of $\sim15-20\%$ in yearly mean GCR flux over a period of one solar cycle in anti-phase with the $\sim$11-year sunspot polar activity cycle is a well-observed known phenomena (see Fig. 1). A number of studies have utilized this change in GCR intensity over solar cycles to suggest (or refute) a possible connection between cosmic rays and climate (clouds, rainfall, temperature etc.). However, significant changes/fluctuations in GCR intensity are observed when the data is averaged over monthly and daily time resolutions. At times, with these time resolutions, the GCR flux is observed to increase/decrease by a large amount (a few percent) during some months in the same year and during several days in the same month. 

The purpose of this paper to search the influence, if any, of GCR flux change on the summer monsoon rainfall in India, at regional and seasonal or even shorter time scales.
  
For this purpose out of 48 years from 1964 - 2011 (see Table 1) we first identify $\sim$one-forth (12) years with lowest rainfall in four Indian summer monsoon months (June-September) (see Table 2). We call them deficient rainfall ('drought') years and the same number of years (12) with the highest rainfall in summer monsoon months (see Table 2). We call them heavy rainfall ('flood') years. In Fig. 2, the deviation of monthly precipitation during superposed ‘drought’ and ‘flood’ years from superposed average precipitation of 48 years (1964-2011) is shown.
 
In Fig. 1, we have plotted yearly average GCR intensity as observed by neutron monitors located at two different latitudes and longitudes (see Table 3), namely Oulu (http://cosmicrays.oulu.fi/) and Newark (http://neutronm.bartol.udel.edu/). Unfortunately, there is no neutron monitor located in India whose data for the period 1964 - 2011 can be utilized for this analysis. However, the time variation shown in Fig. 1 at two locations on the earth is similar in nature globally with different amplitudes at different latitudes. That is, the nature of GCR variations observed at globally distributed monitors are similar, only differing in amplitudes. However, there are some suggestions (e.g. Eroshenko et al., 2010) that the rainfall and the humidity influence the incoming particle flux around the detector; moisture around the detector lowers both the neutrons incident to the surface and albedo neutrons.
 
In Fig. 1, the 'drought' and 'flood' years are indicated by dashed and dotted vertical lines respectively. From Fig. 1 we see that in India droughts/floods can occur at any level of mean GCR flux, minimum/maximum or intermediate level. In other words, these floods/droughts can occur at maximum/minimum/increasing/decreasing phases of the solar activity cycle. Thus if we assume mean GCR flux to be the key, then we can conclude that there is no influence of GCR flux on the Indian monsoon rainfall of inter-annual scale. Next, we proceed to search for any possible influence of GCR flux variability on Indian monsoon rainfall during the same period, assuming that it is more likely that the rainfall changes occur only with GCR flux changes if environmental conditions are suitable, and that not the mean GCR flux, but its variability may affect the rainfall amount/climate (see Laken et al., 2010; Usoskin, 2011).

\begin{figure}[htp!]
   \centerline{\includegraphics[width=\hsize]{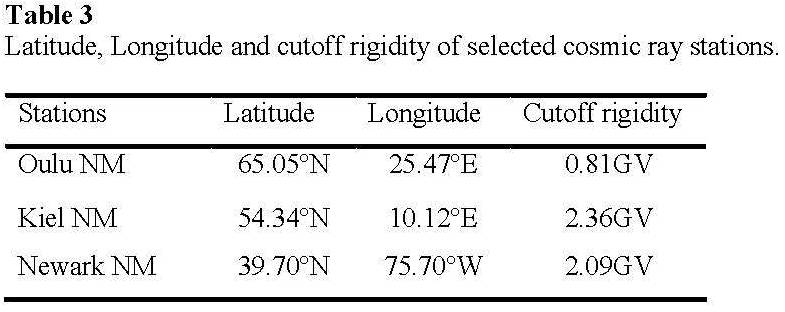}}
        \end{figure}

The cosmic ray count rate has solar cycle dependence, so we normalized the count rate before performing the superposed epoch analysis. Each year's data is normalized to the yearly average for that year, then the data is converted to percentage, which in turn allows for a direct comparison of the different data (i.e. GCR intensity, Sunspot number, 10.7 cm solar radio flux, Total Solar Irradiance). Frist, we have calculated the yearly average for individual years, then each months data is converted into percentage by takeing yearly average as reference. Monthly resolution cosmic ray count rate of Oulu NM during drought and flood years, mean count rate of individual months with standard deviation ($\sigma$) and standard error of mean (SEM = $\frac {\sigma} {\sqrt {n}} $ ) of both before and after normalization are tabulated in Tables 4a and 4b.

We then perform the superposed epoch analysis to study the rate of GCR flux variability during ISMR months (June-September) averaged over 12 drought years and 12 flood years separately. For this purpose we have utilized the normalized GCR intensity data of three neutron monitors located at different positions on the Earth, namely Oulu (Finland), Kiel (Germany) and Newark (USA) (see Table 3). These three location data have been analysed to show that the nature of variation is globally similar, only differing in amplitudes. 

In Fig. 3(a) we have plotted the superposed epoch results of monthly averaged normalized GCR intensity data for deficient rainfall years as observed by Oulu neutron monitor count rate. We see that GCR intensity is decreasing during ISMR (June-September) period (shaded). The rate of decrease has been calculated by fitting a linear curve (see Table 5), taking the pre-monsoon (May) value as a reference. The best-fit result shows that the GCR count rate decreases (negative slope) with linear correlation coefficient R = $-0.95$ (see Table 5 and Fig. 3). 

As the GCR intensity may fluctuate to a large extent on a day-to-day basis, we have done the superposed epoch analysis of the daily normalized GCR count rate, as observed by the Oulu neutron monitor, for the same 12 deficient rainfall years. The result of this analysis is plotted in Fig. 4(a) and tabulated in Table 5. We see a continuously decreasing GCR intensity during the summer monsoon period. We did a linear regression to this averaged data considering the pre monsoon (May) data as the reference. The best-fit line with negative slope (R = $-0.90$) is also shown (see Table 5 and Fig. 4). In both monthly and daily cases, we note that the regression line is steeper than it would be if the regression line is obtained using the entire year.

To show that such a variation is not confined to one location but its nature is global, we did a similar superposed epoch analysis and best fit linear regression, as earlier, for two more neutron monitor stations data namely Kiel and Newark using monthly average GCR count rate as well as daily count rate (see Table 5). We see a similar decreasing trend at these locations also. Thus, we can infer that the trends of rate of change in GCR flux will be similar in nature at Indian locations also.

Next, we consider the same number (12) of heavy rainfall years and did a similar superposed epoch analysis of GCR count rate (both monthly and daily) data for the same three neutron monitors. We also did a linear regression of these data for the ISMR period taking pre-monsoon (May) value as a reference. We find that GCR flux is increasing during ISMR periods. The best-fitted linear curves with positive slope and correlation coefficients are clearly evident on all three-neutron monitor stations and at both time resolutions (see Figs. 3, 4 and Table 5). The linear regression shows a line with positive slope [see Figs. 3(e) and 4(e)], and from these figures it is clear that the slope will be less rapid if a regression line is drawn for the entire year.

\begin{figure*}
   \centerline{\includegraphics[width=\hsize, height =1.1\hsize]{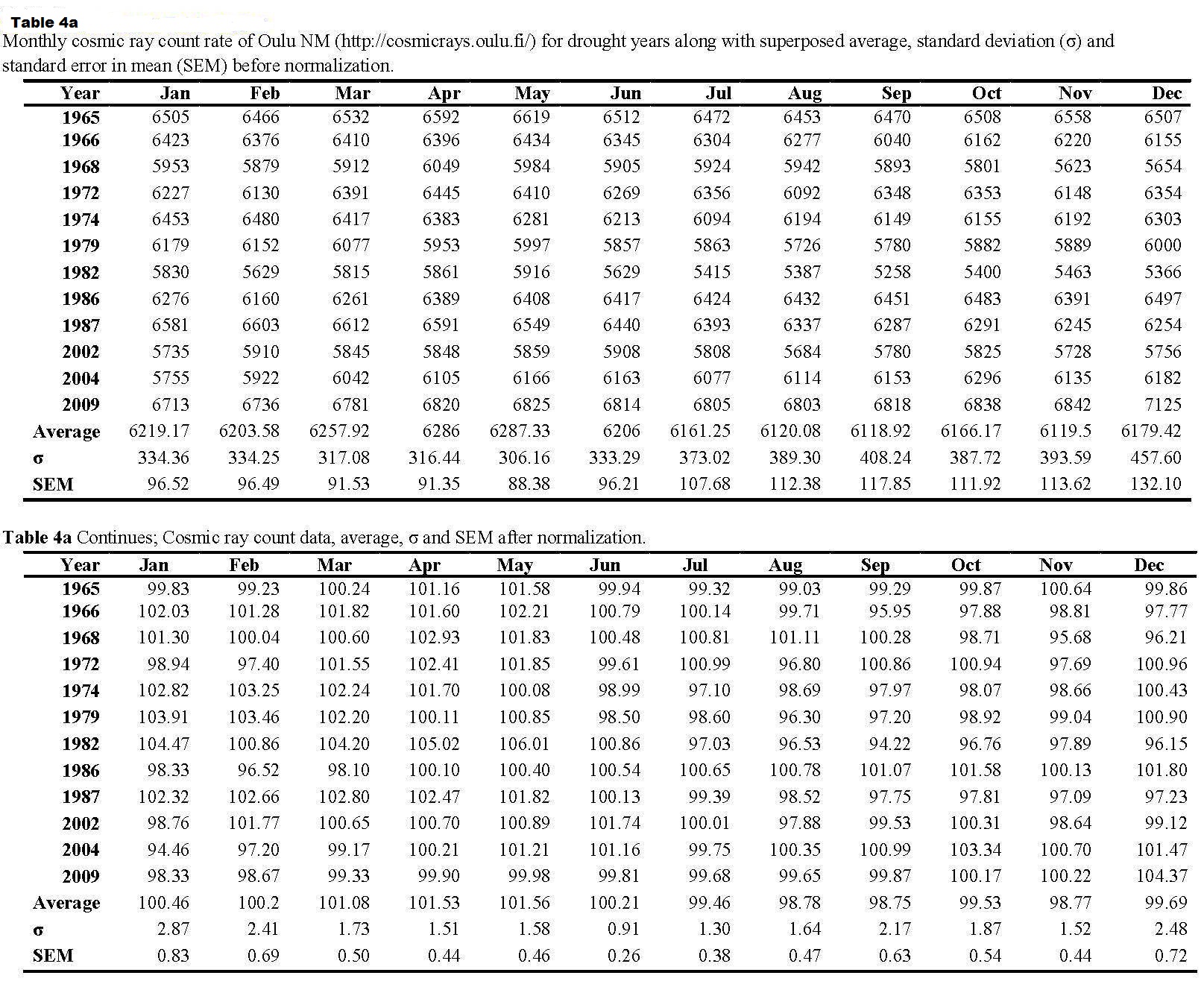}}
        \end{figure*}

In addition to cosmic rays, we extended our analysis to solar activity parameters, such as sunspot number (SSN) and 10.7 cm solar radio flux (http://omniweb.gsfc.nasa.gov/). The SSN is the oldest directly observed solar activity on the photosphere and a very useful indicator of solar activity. The 10.7 cm solar radio flux is an indicator of activity in the upper chromosphere and lower corona. We also considered Total Solar Irradiance (TSI) data; however, this data is available only from 1979 onwards (http://www.ngdc.noaa.gov/; http://lasp.colorado.edu/home/ sorce/data/tsi-data/). 

\begin{figure*}
   \centerline{\includegraphics[width=\hsize, height =1.1\hsize]{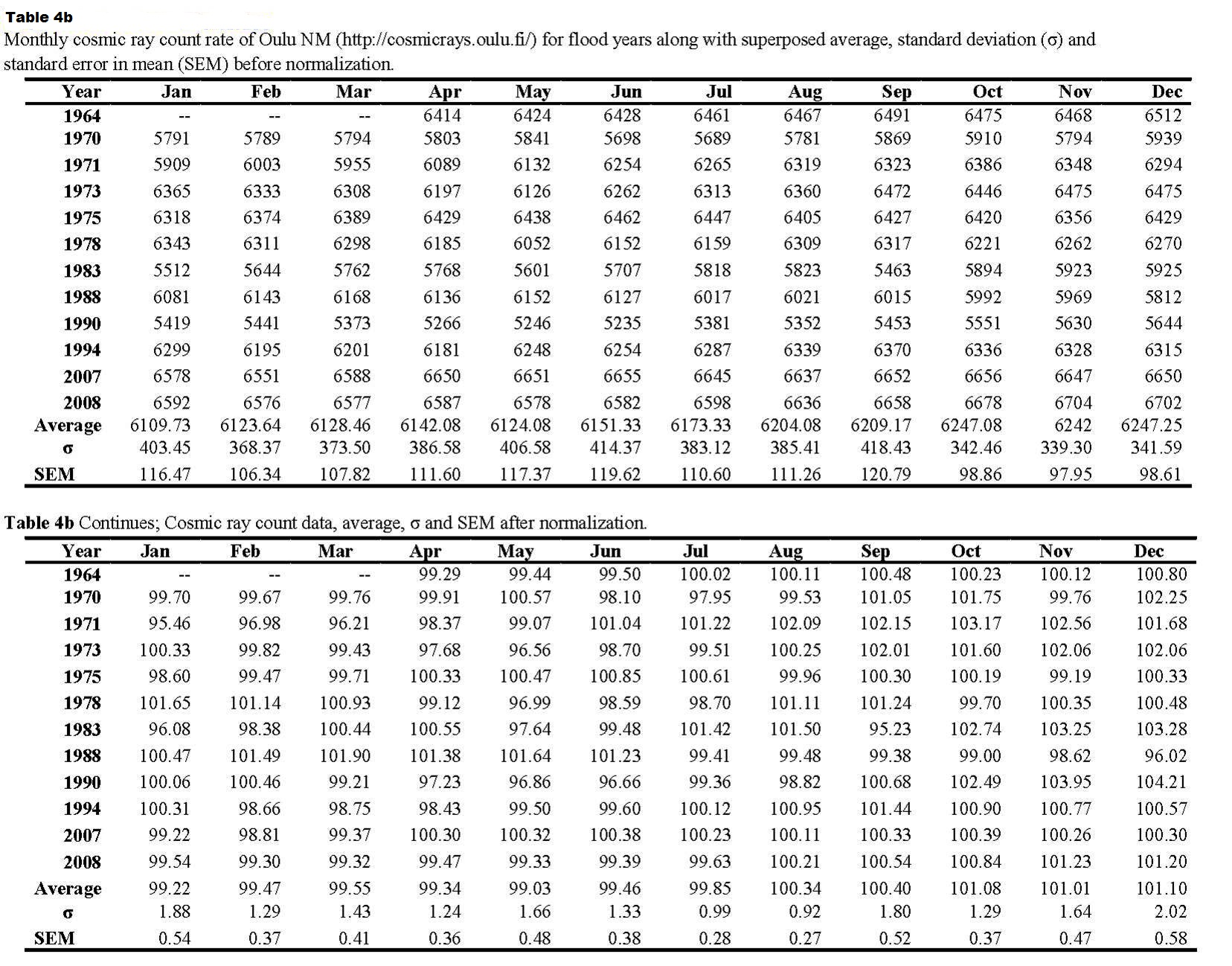}}
        \end{figure*}   

\begin{figure*}
\centerline{\includegraphics[width=\hsize]{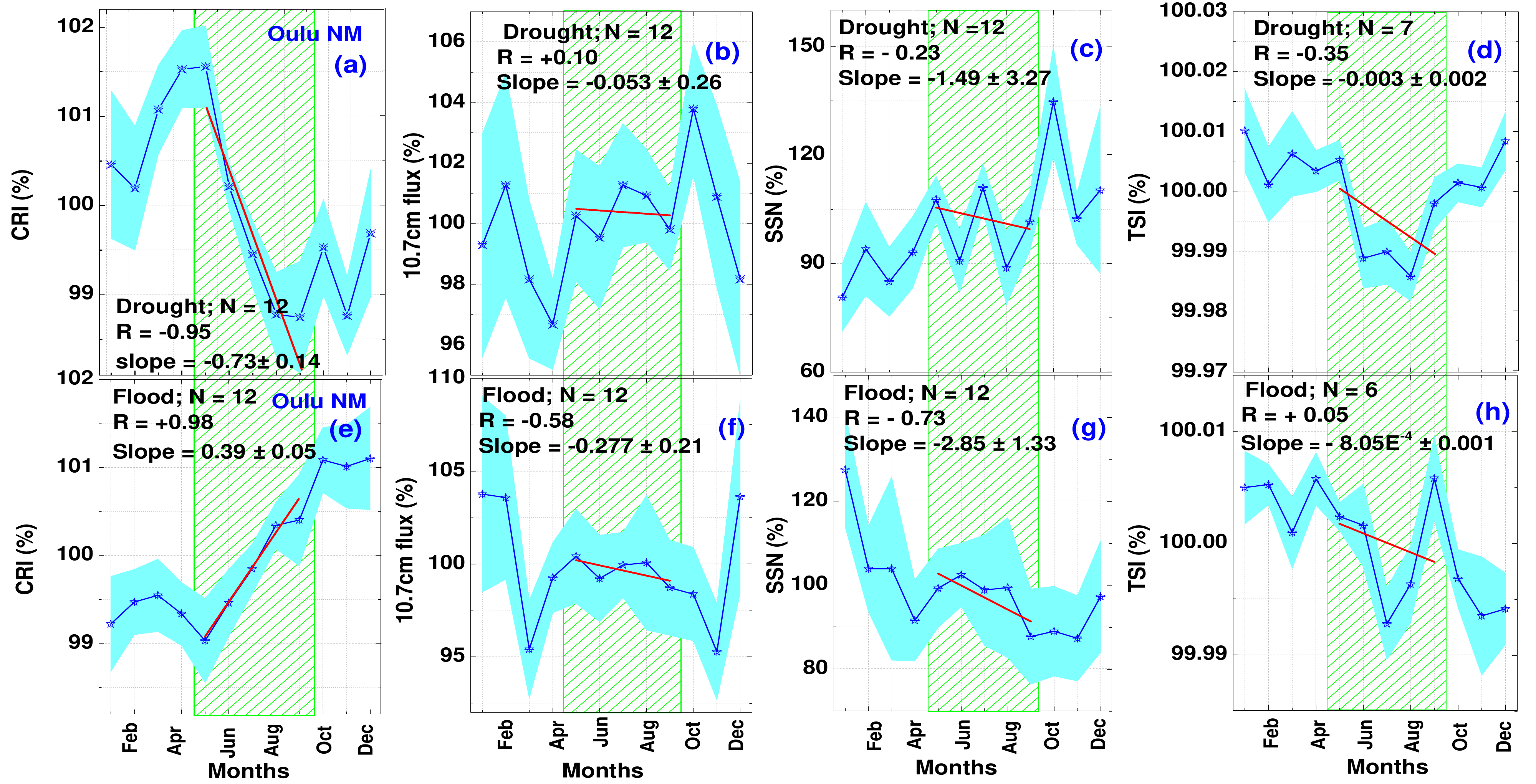}}
\caption{Superposed epoch results of monthly averaged normalized GCR intensity, 10.7cm solar radio flux, SSN and Total solar irradiance along with standard error of mean (color filled around blue line), best-fit linear curve (red straight line) and linear correlation coefficient during ISMR (June-September) period, considering the pre monsoon (May) data as the reference, for deficient rainfall years in upper panel and heavy rainfall years in lower panel.}
\end{figure*}

 \begin{figure*}
\centerline{\includegraphics[width=\hsize]{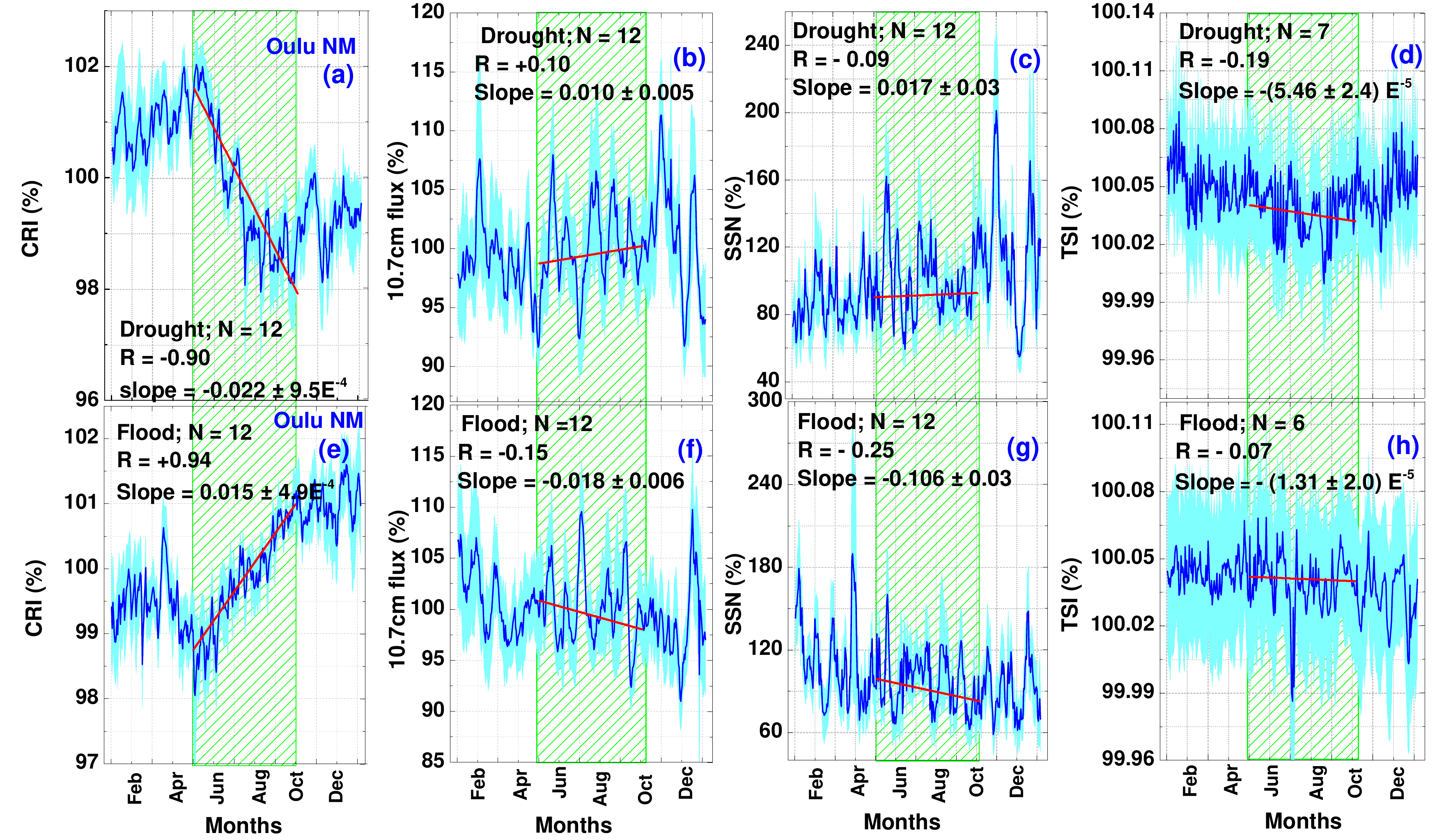}}
\caption{Superposed epoch results of daily averaged normalized GCR intensity, 10.7cm solar radio flux, SSN and Total solar irradiance along with standard error of mean (color filled around blue line), best-fit linear curve (red straight line) and linear correlation coefficient during ISMR (June-September) period, considering the pre monsoon (May) data as the reference, for deficient rainfall years in upper panel and heavy rainfall years in lower panel.}
\end{figure*} 

We analysed both monthly and daily resolution data of solar parameters SSN, 10.7 cm solar radio flux and TSI for drought and flood years. We carried out superposed epoch analysis after normalizing the data as done for GCR count rate. Superposed epoch results of monthly averaged normalized SSN, 10.7 cm solar radio flux and TSI are plotted in Fig. 3 for both drought years [Fig. 3(b-d)] and flood years [Fig. 3(f-h)]. The rate of change has been calculated by fitting a linear curve, taking the pre-monsoon (May) value as a reference. The best-fit results with linear correlation coefficient are also shown (see Fig. 3 and Table 5). We have also done the superposed epoch analysis of the daily normalized solar parameters data, for the same 12 drought ((b-d) of Fig. 4) and flood ((f-h) of Fig. 4) rainfall years. The rate of change has been calculated by fitting a linear curve to the data. The best-fit results with linear correlation coefficient are tabulated in Table 5. We can see the difference in nature of variability (slopes) in GCR flux and solar parameters (SSN, 10.7 cm solar radio flux, TSI), errors in slopes and correlation coefficients (R) during drought and flood periods, favouring GCR flux-rainfall relationship. Noticeable difference seen in GCR variability is not so clear in solar parameters considered here. However, it is possible that restricting the correlation analysis to ISMR months only reduces the apparent dependence on solar indices such as F10.7 and SSN. But the anti-correlation between GCR flux and SSN and F10.7 becomes clearer if the analysis interval is extended to whole year.  

We observe that on an average, the GCR flux is decreasing during ISMR months (June-September) with deficient monsoon rainfall (drought) in India. On the other hand, GCR flux is increasing during ISMR period with heavy rainfall (flood) in India. As regards the change in temperature with rainfall changes, we find that there is a strong inverse relation between the rainfall and temperature (see Fig. 5), at least during ISMR period.

In  the view of the results shown in  Figs. 3 and 4, there is urgent need to quantify the extent of influence, and to identify the physical mechanism(s), responsible for influencing the Indian monsoon rainfall through the cosmic ray flux variability. 
Although, we found definite trends i.e., on the average, heavy rainfall (floods) in India occurs during ISMR period when GCR flux is increasing in the same season, and GCR flux is decreasing in ISMR months during deficient rainfall (drought) in Indian Summer Monsoon period. However, a caveat must be added here; that the rainfall changes can occur with GCR changes only if environmental conditions (to be identified) are suitable. This caveat implies that similar trends in rainfall changes with GCR flux changes (i.e., deficient rainfall associated with decreasing GCR flux and heavy rainfall associated with increasing GCR flux) may not be observed at all geographic locations from equator to pole and in all seasons simultaneously, although the nature of GCR flux change is overall similar at almost all locations on the earth only differing in magnitude. Moreover, there may be exceptions in India even during ISMR season due to unsuitable environmental conditions.

As regard the breakdown of the ENSO-ISMR connection after 1988, mentioned earlier, the breakdown may be because the ISMR is less variable. Similary to ENSO, the GCR flux variability has similar properties on both the quiet interval from 1989 to 2002 and the drought and flood periods before and after this gap. Thus the breakdown in the ISMR-ENSO connection is not necesserly the evidence for the ISMR-GCR hypothesis. The possibility of both the ENSO and GCR variability contributing to ISMR variability in their own way cannot be ruled out at this stage. More efforts and rigorous analyses are required to discriminate between ENSO and GCR as a cause of ISMR variability. 

  \begin{figure}
\centerline{\includegraphics[width=\hsize]{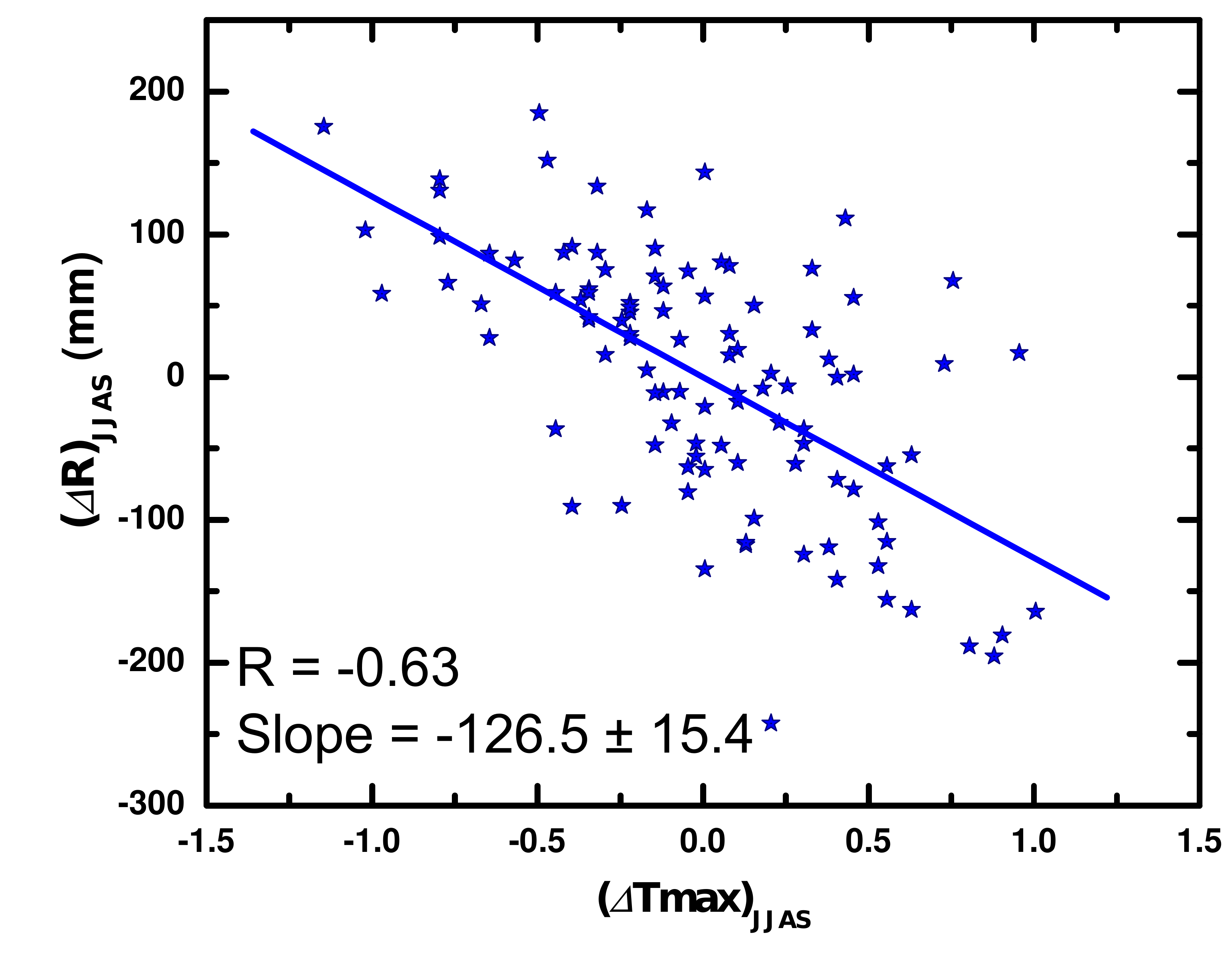}}
\caption{Showing the correlation between deviations from averages in monsoon rainfall [(∆R)JJAS] and maximum temperature [(∆Tmax)JJAS] during ISMR months.}
\end{figure}

\begin{figure*}[htp!]
   \centerline{\includegraphics[width=\hsize]{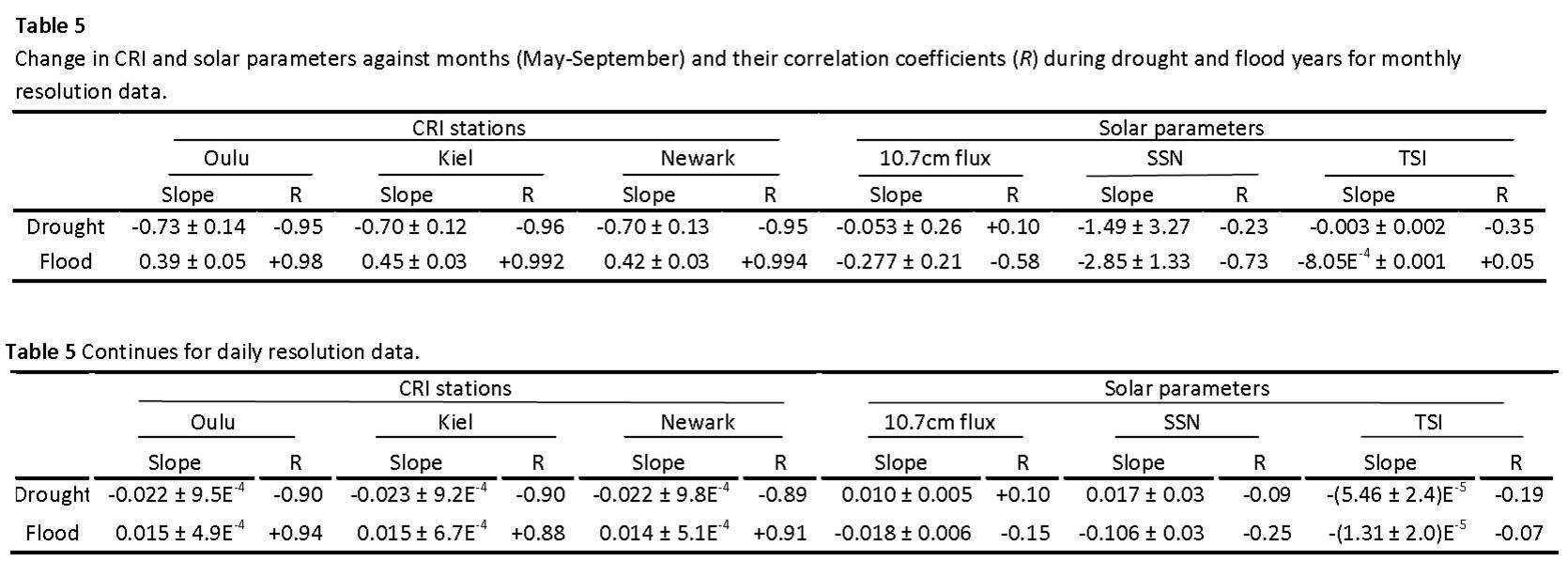}}
        \end{figure*}   

\section{Discussions}
\label{Discussions}
Most of the studies that attempt to study possible GCR-cloud-climate relationship are focused on longer time scales (millennial, centennial, multidecadal and decadal) (e.g. see reviews by Carslaw et al., 2002; Kirkby, 2007; Singh et al., 2011; Rao, 2011; and references therein). However, on shorter time scales too (inter-annual, seasonal and even smaller) attempts have been made to search for this relationship with conflicting results.

Forbush decreases are sudden decreases $\sim$a few percent in cosmic ray intensity within about a day and recover to its pre-reverse level within a week or so (e.g. see Rao, 1972; Venkatesan and Badruddin, 1990; Kudela, 2009 for reviews on cosmic ray variations at different time scales). These Forbush decreases in cosmic rays are thought to be an important laboratory for testing possible cosmic ray climate connection. Decreases in rainfall in the former Soviet Union have been reported in the days of the Forbush decreases (Stozhkov et al., 1995). However, most of the Eastern Mediterranean stations present higher probabilities for a precipitation episode one day after a Forbush decrease (Mavrakis and Lykoudis, 2006). Precipitation changes in relation to GCR flux changes in a short time scale have also been studied by Kniveton and Todd (2001).

A recent claim that Forbush decreases affect atmospheric aerosol and cloud (Svensmark et al., 2009) has been challenged by other studies (Kulmala et al, 2009; Laken et al., 2009; Calogovic et al., 2010) who found no connection between cosmic rays, aerosols and clouds. However, a more recent study (Dragic et al., 2011) from an analysis of European region data supports the idea that cosmic rays influence the atmospheric process and climate. Earlier too, the claim of decreased cloudiness detected during Forbush decreases (Pudovkin and Veretenenko, 1995) was not observed by Palle and Butler (2001) even during the same Forbush decreases. Moreover, the underlying physics suggesting for a connection between cosmic rays, aerosols and cloud is still highly speculative (Legras et al., 2010) and empirical evidences for cosmic ray-cloud relation is still inconclusive (Usoskin, 2011).

It was concluded in a review (Kirkby, 2007), based on the available results for longer timescales (millennial, centennial and multi decadal), that increased GCR flux appears to be associated with a cooler climate and a weakening of the monsoon; and a decreased GCR flux is associated with a warmer climate and strengthening of the monsoon. From our observations on a much shorter time scale during monsoon season in India, we observe that a decreasing GCR flux corresponds to decreasing rainfall and increasing GCR flux corresponds to increasing rainfall. Moreover, our preliminary results reported earlier (Badruddin et al., 2006, 2009) show that temperature and rainfall changes show an opposite behaviour, i.e., temperature is enhanced during deficient ISMR periods and it is lower in heavy rainfall ISMR periods. 

The observation that the cosmic ray intensity is decreasing during ISMR months in almost all the years which, are deficient in rainfall ('drought' years) may be interpreted to suggest that a GCR -rainfall relation is possible in Indian sub-continent during ISMR periods, at least. Thus, the GCR-rainfall relation should be considered as  a potentially important driver of climate variability.

A significant part of precipitation that falls in the tropics is warm rain formed by coalescence of cloud droplets (Kostinski and Shaw, 2005). Formation of cloud droplets requires a water vapour super saturation environment and particles able to act as cloud condensation nuclei. Usually cloud droplets are formed on aerosol particles containing a certain stable fraction. After condensation droplets grow by vapour diffusion and droplet-droplet collision (coalescence), the latter providing more rapid growth as droplet size increases (Harrison and Ambaum, 2009). Electrical effects play an important role in cloud microphysics. Both condensation and coalescence can be influenced by the charge (Pruppacher and Klett, 1997; Tinsley, 2008). Early laboratory studies found that raindrops (of around 0.5mm diameter) are about factor 100 more efficient at collecting aerosols when they are charged rather that neutral (Barlow and Latham, 1983). Grover and Beard (1975) calculated collision efficiencies and found a significant increase in collision efficiency when the droplets were loaded with a charge of the magnitude typical of thunderstorm clouds. Khain et al. (2004) from their simulation results have shown that the injection of just a small fraction of charged particles rapidly triggered the collision process and lead to raindrop formation a few minutes after the injection, thus seeding with charged particles may be a very efficient tool for rain enhancement, they suggested. The collision efficiencies highly depend on droplet charge and size. The collision efficiency is much enhanced in the case of a charged droplet collisions than in case of neutral droplet collisions. More specifically, they found that the collision efficiency between charged and neutral droplets, as well as between droplet charges of opposite polarity, is many orders higher than in the case of gravity-induced collisions. Thus, efficient collision takes place between cloud droplets and coalescence to large droplet is enhanced by electrical forces between charged droplets. This significantly increases the rate of raindrop formation (Khain et al., 2004). Another potential mechanism may operate through electrostatic image forces. Because of electrostatic image forces, electrical forces between charged droplets are always attractive at small separations whatever the relative polarities of the colliding particles (Tinsley, 2008). In this case, the attraction between droplets may lead to droplet size increase. As the droplet size increases, the droplet-droplet collision (coalescence) will lead to more rapid growth, leading to enhanced rate of raindrop formation, as suggested by Harrison and Ambaum (2009). This mechanism appears more likely as the rate of change of GCR flux and not the amount of GCR flux is considered to be the key factor. However, more simulations and experiments need to be performed to demonstrate clearly how the increasing GCR flux corresponds to increasing rainfall and vice versa.

Thus it is expected that in proper atmospheric/environmental conditions (e.g. air humidity, aerosols, temperature, cloud type etc.) increasing GCR flux will increase coalescence efficiency that will lead to bigger rain droplets while decreasing GCR flux will decrease the coalescence efficiency and will suppress the droplet growth. During decreasing flux of cosmic rays, levitation/dispersion of low clouds due to electrical effects (Levin and Ziv, 1974) may also play some role in such a way that it disperses the low cloud amount in proper climatic conditions. We suspect that in suitable environmental conditions, charge particle (cosmic ray) flux rate change modulates the droplet collision and coalescence efficiency and affects the rainfall to certain extent.

Rain formation is a function of different parameters of macro- and micro-physics. The important parameter for the microphysics is the ambient temperature where clouds reside and formation of raindrops occurs due to process of spontaneous coalescence and accretion (Rogers and Yau, 1989). Although the initiation of raindrop coalescence remain an unsolved problem in cloud physics (Kostinski and Shaw, 2005), we suspect that charge induced cloud microphysics, for example, accelerating/decelerating coalescence to larger raindrops (Harrison and Ambaum, 2009) is the likely effect that  plays some role in affecting the rainfall variability in India during Indian Summer Monsoon Season, at least, depending on the increasing/decreasing rate of change of charge particle (cosmic ray) flux in the corresponding period, under suitable environmental conditions (e.g. cloud type, temperature, pressure etc.).

Monsoon rainfall variability is connected with global precipitation (Hulme et al., 1998). There is a strong inverse relationship between the monsoon variability and tropical belt temperature (see Dugam and Kakade, 1999; Badruddin et al., 2006, 2009). Considering that change in monsoon rainfall variability is also consistent with the change in global mean precipitation (Hulme et al., 1998) and precipitation/rainfall is inversely related to temperature (see Fig. 8), we suspect that the monsoon rainfall variability may have some influence on the changes in global temperature also. Thus, it should be clarified whether monsoon/rainfall variability plays any role in global warming or its effects are only local. It has been suggested (Ban-Weiss et al., 2011) that evaporated water helps in cooling earth as a whole and not just the local area of evaporation. On the other hand, reduction in evaporated water is likely to contribute to global warming significantly.

The possible influence of GCR on clouds is a controversial issue. It appears that GCR flux variability plays an important role in influencing the ISMR in this season, at least. It is likely, as we suspect, that the physical state of the cloud droplets may play a significant role. Local physical (cloud type, temperature, humidity etc.) and chemical conditions may play a major role (Enghoff et al., 2011; Duplissy et al., 2010; Kirkby et al., 2011) in deciding the extent of the influence. Physics of liquid and ice cloud may differ (Geirens and Ponater, 1999). Low clouds generally consist of liquid water droplets (Marsh and Svensmark, 2000). It has been suggested that low cloud liquid droplets over the tropics are more sensitive to cosmic ray variability (Palle and Butler, 2000). It is suggested that such studies (i.e., effects of GCR flux variability on rainfall variability) on regional hydrological regions have to be studied in more detail. It is also suggested that proper environmental condition in which the influence of GCR flux variability of rainfall is more significant, needs to be identified.

Although amount of cloud may be dependent on GCR flux, in our hypothesis, we do not consider a direct relationship between the GCR flux and amount of cloud cover as the key; it is still controversial. We propose an alternate scenario, although speculative, in which the cosmic ray variability influences the rainfall from clouds that are formed in proper environmental conditions. We propose that increasing/decreasing GCR flux influences the rainfall which later results in enhanced/reduced evaporation. This change in evaporation from the Earth surface influences the low cloud amount which in turn alters the planetary albedo and concequently there is change in the temperature. However, such influence is only regional or has global effect needs to verified. Therefore, more research is needed to understand the relationship amoung variability in GCR, ISMR, surface evaporation, low cloud, planetary albedo and temperature. Also model studies are needed to understand the extent to which such variability influences the regional and global rainfall and temperature. 
   
\section{Conclusions}
\label{Conclusions}
We find that the decreasing cosmic ray flux does play a role in such a way that the rainfall over this region of the globe (India), at least, is reduced when cosmic ray flux is decreasing. We speculate that the hypothesis, proposed here, on the basis of Indian climate data, can be extended to whole tropical and sub-tropical belt, and that it may contribute to global temperature in some way.

In conclusion, a GCR-ISMR link seems plausible and the GCR-rainfall relation should be considered as a potentially important driver of climate variability. However, further studies are required to improve our understanding of the link between cosmic rays and summer monsoon climate over India. It is also required to fully investigate the contributions of possible mechanisms, discussed here, to the variability in precipitation. Further, once our hypothesis is confirmed, there is an urgent need to identify the local physical and chemical conditions conducive for significant effect of GCR flux variability in influencing the rainfall/precipitation.

We suggest the following scenario, although speculative, for possible relationship between GCR flux-rainfall-temperature. 

\begin{enumerate}

 \item Increasing GCR flux $\longrightarrow$ increasing rainfall $\longrightarrow$ enhanced surface evaporation $\longrightarrow$ increased low cloud $\longrightarrow$ more scattering of solar radiation back to space (more planetary albedo) $\longrightarrow$ lower temperature.
\item Decreasing GCR flux $\longrightarrow$ decreasing rainfall $\longrightarrow$ decreased surface evaporation $\longrightarrow$ reduced low cloud $\longrightarrow$ less scattering of solar radiation back to space (less planetary albedo) $\longrightarrow$ higher temperature.

   \end{enumerate}

 \section*{Acknowledgements}
     We thank Station Manager Ilya Usoskin and Sodankyla Geophysical Observatory for the online availability of Oulu neutron monitor data, The National Science Foundation (supporting Bartol Research Institute neutron monitors) and Principle Investigator John W. Bieber for the online availability of Newark neutron monitor data and Christian T. Steigies and Extraterrestrial Physics Department of University of Kiel for the online availability of Kiel neutron monitor data. Availability of Indian climate data through Indian Institute of Tropical Meteorology Pune's website and its use is gratefully acknowledged with thanks. We also acknowledge the use of SSN and 10.7 cm solar radio flux data available through the NASA/GSFC OMNI Web interface, Total Solar Irradiance data through National Geophysical Data Center website and SORCE homepage. The authors also thank the Editor and Referees, whose comments and suggestions helped us to improve the paper.


\begin{thebibliography}{00}

\bibitem[\protect\citeauthoryear{Agnihotri et al.(2002)}{Agnihotri et al., 2002}]{Agnihotri2002}
Agnihotri, R., Dutta, K., Bhushan, R., Somayajulu, B. L. K., 2002. Evidence for solar forcing on the Indian monsoon during the last millennium. Earth Planet. Sci. Lett. 198, 521-527.

\bibitem[\protect\citeauthoryear{Ashok et al.(2001)}{Ashok et al., 2001}]{Ashok2001}
Ashok, K., Guan, Z., Yamagata, T., 2001. Impact of Indian Ocean Dipole on the relationship between the Indian monsoon rainfall and ENSO. Geophys. Res. Lett. 28, 4499-4502.

\bibitem[\protect\citeauthoryear{Badruddin et al.(2006)}{Badruddin et al., 2006}]{Badruddin2006}
Badruddin, Singh, Y. P., Singh, M., 2006. Does solar variability affect Indian (Tropical) weather and climate?: An assessment. In: Gopalswamy, N., Bhattacharya, A. (Eds.), Solar influence on the Heliosphere and Earth’s Environment: Recent Progress and Prospects (Proc. ILWS Workshop). Quest Publications, 444-447.

\bibitem[\protect\citeauthoryear{Badruddin et al.(2009)}{Badruddin et al., 2009}]{Badruddin2009}
Badruddin, Aslam, O. P. M., Singh, M., 2009. Influence of solar and cosmic-ray variablity on climate. Proc. 31st Int. Cosmic Ray Conf. Lodz, SH 3.4, 1-3.

\bibitem[\protect\citeauthoryear{Ban-Weiss et al.(2011)}{Ban-Weiss et al., 2011}]{Ban-Weiss2011}
Ban-Weiss, G. A., Bala, Govindaswamy, Cao, L., Pongaratez, J., Caldeera, K., 2011. Climate forcing and response to idealized changes in surface latent and sensible heat. Environ. Res. Lett. 6, 034032.

\bibitem[\protect\citeauthoryear{Bazilevskaya and Svirzhevskaya (1998)}{Bazilevskaya and Svirzhevskaya, 1998}]{Bazilevskaya1998}
Bazilevskaya, G. A., Svirzhevskaya, A. K., 1998. On the stratospheric measurements of cosmic rays. Space Sci. Rev. 85, 431–521.

\bibitem[\protect\citeauthoryear{Barlow and Latham (1983)}{Barlow and Latham, 1983}]{Barlow1983}
Barlow, A. K., Latham, J., 1983. A laboratory study of the scavenging of sub-micron aerosol by charged raindrops. Royal Meteorological Society, Quart. J. 109, 763-770. 

\bibitem[\protect\citeauthoryear{Bhalme et al.(1981)}{Bhalme et al., 1981}]{Bhalme1981}
Bhalme, H. N., Reddy, R. S., Mooley, D. A., Ramana Murty, Bh. V., 1981.  Solar activity and Indian weather/climate.  Earth Planet. Sci. 90, 245-262.

\bibitem[\protect\citeauthoryear{Bhattacharya and Narasimha (2005)}{Bhattacharya and Narasimha, 2005}]{Bhattacharya2005}
Bhattacharya, S., Narasimha, R., 2005. Possible association between Indian monsoon rainfall and solar activity. Geophys. Res. Lett. 32, L05813.

\bibitem[\protect\citeauthoryear{Calogovic et al.(2010)}{Calogovic et al., 2010}]{Calogovic2010}
Calogovic, J., Albert, C., Arnold, F., Beer, J., Desorgher, L., Flueckiger, E. O., 2010. Sudden cosmic ray decreases: No change of global cloud cover. Geophys. Res. Lett. 37, L03802.

\bibitem[\protect\citeauthoryear{Carslaw et al.(2002)}{Carslaw et al., 2002}]{Carslaw2002}
Carslaw, K. S., Harrison, R. G., Kirkby, J., 2002. Cosmic Rays, Clouds, and Climate. Science, 298, 1732-1737.


\bibitem[\protect\citeauthoryear{Dragić et al.(2011)}{Dragić et al., 2011}]{Dragić2011}
Dragić, A., Aničin, I., Banjanac, R., Udovičić, V., Joković, D., Maletić, D., Puzović, J., 2011. Forbush decreases-clouds relation in the neutron monitor era. Astrophys. Space Sci. Trans. 7, 315-318.



\bibitem[\protect\citeauthoryear{Dugam and Kakade (1999)}{Dugam and Kakade, 1999}]{Dugam1999}
Dugam, S. S., Kakade, S. B., 1999. Global temperature and monsoon activity. Proc. Indian Acad. Sci. (Earth and Planet. Sci.), 108, 305-307. 

\bibitem[\protect\citeauthoryear{Duplissy et al.(2010)}{Duplissy et al., 2010}]{Duplissy2010}
Duplissy, J., Enghoff, M. B., Aplin, K. L., Arnold, F., et al., 2010. Results from the CERN pilot CLOUD experiment. Atmos. Chem. Phys. 10, 1635-1647.

\bibitem[\protect\citeauthoryear{Enghoff et al.(2011)}{Enghoff et al., 2011}]{Enghoff2011}
Enghoff, M. B., Pedersen, J. O. P., Uggerhoj, U. I., Paling, S. M., Svensmark, H., 2011. Aerosol nucleation induced by a high energy particle beam. Geophys. Res. Lett. 38, L09805. 

\bibitem[\protect\citeauthoryear{Eroshenko et al.(2010)}{Eroshenko et al., 2010}]{Eroshenko2010}
Eroshenko, E., Velinov, P., Belov, A., Yanke, V., Pletnikov, E., Tassev, Y., Mishev, A., Mateev, L., 2010. Relationships between neutron fluxes and rain flows. Adv. Space Res. 46, 637-641.  

\bibitem[\protect\citeauthoryear{Gadgil et al.(2004)}{Gadgil et al., 2004}]{Gadgil2004}
Gadgil, S., Vinayachandran, P. N., Francis, P. A., Gadgil, S., 2004. Extremes of the Indian summer monsoon rainfall, ENSO and equatorial Indian Ocean oscillation. Geophys. Res. Lett. 31, L12213.


\bibitem[\protect\citeauthoryear{Gierens and Ponater (1999)}{Gierens and Ponater, 1999}]{Gierens1999}
Gierens, K., Ponater, M., 1999. Comment on 'Variation of cosmic ray flux and global cloud coverage - a missing link in solar-climate relationships' by H. Svensmark and E. Friis-Christensen (1997).  J. Atmos. Sol. Terr. Phys. 61, 795- 797. 

\bibitem[\protect\citeauthoryear{Grover and Beard (1975)}{Grover and Beard, 1975}]{Grover1975}
Grover, S. N., Beard, K. V., 1975. A numerical determination of the efficiency with which electrically charged cloud drops and small raindrops collide with electrically charged spherical particles of various densities. J. Atmos. Sci. 32, 2156-2165.

\bibitem[\protect\citeauthoryear{Gupta et al.(2005)}{Gupta et al., 2005}]{Gupta2005}
Gupta, A. K., Das, M., Anderson, D. M., 2005. Solar influence on the Indian summer monsoon during the Holocene. Geophys. Res. Lett.  32, L17703.

\bibitem[\protect\citeauthoryear{Harrison and Ambaum (2009)}{Harrison and Ambaum, 2009}]{Harrison2009}
Harrison, R. G., Ambaum, M. H. P., 2009. Observed atmospheric electricity effect on clouds. Environ. Res. Lett. 4, 014003.

\bibitem[\protect\citeauthoryear{Hiremath and Mandi (2004)}{Hiremath and Mandi, 2004}]{Hiremath2004}
Hiremath, K. M., Mandi, P. I., 2004. Influence of the solar activity on the Indian Monsoon rainfall. New Astron. 9, 651-662.

\bibitem[\protect\citeauthoryear{Hong et al.(2001)}{Hong et al., 2001}]{Hong2001}
Hong, Y.T., Wang, Z. G., Jiang, H. B., Lin, Q. H., Hong, B., Zhu, Y. X., Wang, Y., Xu, L. S., Leng, X. T., Li, H. D., 2001. A 6000-year record of changes in drought and precipitation in northeastern China based on a δ13C time series from peat cellulose. Earth Planet. Sci. Lett. 185, 111-119.

\bibitem[\protect\citeauthoryear{Hulme et al.(1998)}{Hulme et al., 1998}]{Hulme1998}
Hulme, M., Timothy O. J.,  Timothy J. C., 1998. Precipitation sensitivity to global warming: Comparison of observations with HadCM2 simulations. Geophys. Res. Lett. 25, 3379-3382. 

\bibitem[\protect\citeauthoryear{Jagannathan and Bhalme (1973)}{Jagannathan and Bhalme, 1973}]{Jagannathan1973}
Jagannathan, P., Bhalme, H. N., 1973. Changes in the Pattern of Distribution of Southwest Monsoon Rainfall Over India Associated With Sunspots . Mon. Weather Rev. 101, 691-700.

\bibitem[\protect\citeauthoryear{Khare and Nigam (2006)}{Khare and Nigam, 2006}]{Khare2006}
Khare, N., Nigam, R., 2006. Can the possibility of some linkage of monsoonal precipitation with solar variability be ignored? Indications from foraminiferal proxy records. Curr. Sci., 90, 1685-1688.

\bibitem[\protect\citeauthoryear{Khain et al.(2004)}{Khain et al., 2004}]{Khain2004}
Khain, A., Pokrovsky, A., Pinsky, M., Seifert, A., Phillips, V., 2004. Simulation of Effects of Atmospheric Aerosols on Deep Turbulent Convective Clouds Using a Spectral Microphysics Mixed-Phase Cumulus Cloud Model. Part I: Model Description and Possible Applications. J. Atmos. Sci. 61, 2963-2982.

\bibitem[\protect\citeauthoryear{Kirkby (2007)}{Kirkby, 2006}]{Kirkby2007}
Kirkby, J., 2007. Cosmic rays and climate. Surv. Geophys. 28, 333-375.

\bibitem[\protect\citeauthoryear{Kirkby et al.(2011)}{Kirkby et al., 2011}]{Kirkby2011}
Kirkby, J., Curtius, J., Almeida, J., Dunne, E., et al., 2011. Role of sulphuric acid, ammonia and galactic cosmic rays in atmospheric aerosol nucleation. Nature, 476, 429-433.

\bibitem[\protect\citeauthoryear{Kniveton and Todd (2001)}{Kniveton and Todd, 2001}]{Kniveton2001}
Kniveton, D. R., Todd, M. C., 2001. On the relationship of cosmic ray flux and precipitation.  Geophys. Res. Lett. 28, 1527-1530.

\bibitem[\protect\citeauthoryear{Kostinski and Shaw (2005)}{Kostinski and Shaw, 2005}]{Kostinski2005}
Kostinski, A. B., Shaw, R. A., 2005. Fluctuations and Luck in Droplet Growth by Coalescence. Bulletin of the American Meteorological Society, 86, 235-244. 

\bibitem[\protect\citeauthoryear{Kripalani and Kulkarni (1997)}{Kripalani and Kulkarni, 1997}]{Kripalani1997}
Kripalani, R. H., Kulkarni, A., 1997. Climate impact of EI Niño/La Niña on the Indian monsoon: A new perspective. Weather, 52, 39-46.  

\bibitem[\protect\citeauthoryear{Kripalani et al. (2003)}{Kripalani et al., 2003}]{Kripalani2003}
Kripalani, R. H., Kulkarni, A., Sabade, S. S., Khandekar, M. L., 2003. Indian Monsoon Variability in a Global Warming  Scenario. Natural 
Hazards, 29, 189-206.

\bibitem[\protect\citeauthoryear{Kudela (2009)}{Kudela, 2009}]{Kudela2009}
Kudela, K., 2009. On energetic particles in space. Acta Phys. Slovaca, 59, 537-652. 

\bibitem[\protect\citeauthoryear{Kulmala et al. (2009)}{Kulmala et al., 2009}]{Kulmala2009}
Kulmala, M., Asmi, A., Lappalainen, H. K., Carslaw, K. H., et al., 2009. Introduction: European Integrated Project on Aerosol Cloud Climate and Air Quality interactions (EUCAARI) - integrating aerosol research from nano to global scales. Atmos. Chem. Phys. 9, 2825-2841. 

\bibitem[\protect\citeauthoryear{Kumar et al. (1999)}{Kumar et al., 1999}]{Kumar1999}
Kumar K. K., Rajagopalan,  B., Cane, A., 1999. On the weakening relationship between the Indian monsoon and ENSO. Science, 284, 2156-2159. 

\bibitem[\protect\citeauthoryear{Kumar et al. (2002)}{Kumar et al., 2002}]{Kumar2002}
Kumar, R. R.,  Kumar, K. K.,  Ashrit, R. G., Patwardhan, S. K., Pant, G. B., 2002. Climate Change in India. Shukla, J., et al., (Ed.), Tata McGraw Hill, New Delhi, India, 24–75.


\bibitem[\protect\citeauthoryear{Laken et al. (2009)}{Laken et al., 2009}]{Laken2009}
Laken, B., Wolfendale, A., Kniveton, D., 2009. Cosmic ray decreases and changes in the liquid water cloud fraction over the oceans. Geophys. Res. Lett. 36, L23803. 

\bibitem[\protect\citeauthoryear{Laken et al. (2010)}{Laken et al., 2010}]{Laken2010}
Laken, B. A., Kniveton, D. R., Frogley, M. R., 2010. Cosmic rays linked to rapid mid-latitude cloud changes. Atmos. Chem. Phys. 10, 10941-10948.

\bibitem[\protect\citeauthoryear{Legras et al. (2010)}{Legras et al., 2010}]{Legras2010}
Legras, B., Mestre, O., Brad, E., Yiou, P., 2010. A critical look at solar-climate relationships from long temperature series.  Climate of the Past, 6, 745-758.

\bibitem[\protect\citeauthoryear{Levin and Ziv (1974)}{Levin and Ziv, 1974}]{Levin1974}
Levin, Z., Ziv, A., 1974. The electrification of thunderclouds and the rain gush.  J. Geophy. Res. 79, 2699.

\bibitem[\protect\citeauthoryear{Marsh and Svensmark (2000)}{Marsh and Svensmark, 2000}]{Marsh2000}
Marsh, N. D., Svensmark, H., 2000. Low Cloud Properties Influenced by Cosmic Rays.  Phys. Rev. Lett. 85, 5004-5007. 

\bibitem[\protect\citeauthoryear{Mavrakis and Lykoudis (2006)}{Mavrakis and Lykoudis, 2006}]{Mavrakis2006}
Mavrakis, A., Lykoudis, S., 2006. Heavy precipitation episodes and cosmic rays variation. Adv. Geosci. 7, 157-161.

\bibitem[\protect\citeauthoryear{Neff et al. (2001)}{Neff et al., 2001}]{Neff2001}
Neff, U., Burns, S. J., Mangini, A., Mudelsee, M., Fleitmann, D., Matter, A., 2001. Strong coherence between solar variability and the monsoon in Oman between 9 and 6kyr ago.  Nature, 411, 290-293. 

\bibitem[\protect\citeauthoryear{Palle and Butler (2000)}{Palle and Butler, 2000}]{Palle2000}
 Palle, B. E., Butler, C. J., 2000. Cosmic rays and climate: The influence of cosmic rays on terrestrial clouds and global warming. Astron. Geophys. 41, 4.18-4.22.		

\bibitem[\protect\citeauthoryear{Pant and Parthasarathy (1981)}{Pant and Parthasarathy, 1981}]{Pant1981}
Pant, G. B., Parthasarathy, B., 1981. Some aspects of an association between the southern oscillation and Indian summer monsoon. Arch. Meteorol. Geophys. Bioklimatol. Ser. B, 29, 245–251.


\bibitem[\protect\citeauthoryear{Parker (1999)}{Parker, 1999}]{Parker1999}
Parker, E. N., 1999. Solar physics: Sunny side of global warming. Nature, 399, 416-417.



\bibitem[\protect\citeauthoryear{Pruppacher and Klett (1997)}{Pruppacher and Klett, 1997}]{Pruppacher1997}
Pruppacher, H. R., Klett, J. D., 1997. Microphysics of Clouds and Precipitation, (Second Revised and Enlarged Edition with an Introduction to Cloud Chemistry and Cloud Electricity), Kluwer Academic Publishers, Dordrecht, pp 954.

\bibitem[\protect\citeauthoryear{Pudovkin and Veretenenko (1995)}{Pudovkin and Veretenenko, 1995}]{Pudovkin1995A}
Pudovkin, M. I., Veretenenko, S.V., 1995.  Cloudiness decreases associated with Forbush decreases of galactic cosmic rays. J. Atmos. Terr. Phys. 57, 1349-1355.


\bibitem[\protect\citeauthoryear{Rao (1972)}{Rao, 1972}]{Rao1972}
Rao, U. R, 1972. Solar modulation of galactic cosmic radiation. Space Sci. Rev. 12, 719-809.  

\bibitem[\protect\citeauthoryear{Rao (2011)}{Rao, 2011}]{Rao2011}
Rao, U. R., 2011. Contribution of changing galactic cosmic ray flux to global warming. Current Science, 100, 223-225. 

\bibitem[\protect\citeauthoryear{Rogers and Yau (1989)}{Rogers and Yau, 1989}]{Rogers1989}
Rogers, R. R., Yau, M. K., 1989. A short course in cloud physics, Pergamon Press, Oxford.

\bibitem[\protect\citeauthoryear{Ruzmaikin et al. (2006)}{Ruzmaikin et al., 2006}]{Ruzmaikin2006}
Ruzmaikin, A., Feynman, J., Yung, Y. L., 2006. Is solar variability reflected in the Nile River?  J. Geophys. Res. 111, D21114.

\bibitem[\protect\citeauthoryear{Rasmusson and Carpenter (1983)}{Rasmusson and Carpenter, 1983}]{Rasmusson1983}
Rasmusson, E. M., Carpenter, T. H., 1983. The relationship between eastern equatorial Pacific sea surface temperatures and rainfall over India and Sri Lanka. Mon. Weather Rev. 111, 517–528. 

\bibitem[\protect\citeauthoryear{Shukla (2007)}{Shukla, 2007}]{Shukla2007}
Shukla, J., 2007. Monsoon Mysteries. Science, 318, 204-205. 

\bibitem[\protect\citeauthoryear{Sikha (1980)}{Sikha, 1980}]{sikha1980}
Sikka, D. R., 1980. Some aspects of the large-scale fluctuations of summer monsoon rainfall over India in relation to fluctuations in the planetary regional scale circulation parameters. Proc. Indian Acad. Sci. Earth Planet. Sci. 89, 179-195.

\bibitem[\protect\citeauthoryear{Singh et al. (2011)}{Singh et al., 2011}]{Singh2011}
Singh, A. K., Siingh, D., Singh, R. P., 2011.  Impact of galactic cosmic rays on Earth’s atmosphere and human health . Atmos. Environ. 45, 3806-3818. 

\bibitem[\protect\citeauthoryear{Singh and Badruddin (2006)}{Singh and Badruddin, 2006}]{Singh2006}
Singh, Y. P., Badruddin, 2006. Statistical considerations in superposed epoch analysis and its applications in space research. J. Atmos. Solar-Terr. Phys. 68, 803-813.

\bibitem[\protect\citeauthoryear{Sinha et al. (2007)}{Sinha et al., 2007}]{Sinha2007}
Sinha, A., Cannariato, K. G., Stott, L. D., Cheng, H., Edwards, R. L., Yadava, M. G., Ramesh, R., Singh, I. B., 2007. A 900-year (600 to 1500 A.D.) record of the Indian summer monsoon precipitation from the core monsoon zone of India. Geophys. Res. Lett. 34, L16707. 


\bibitem[\protect\citeauthoryear{Stozhkov et al. (1995)}{Stozhkov et al., 1995}]{Stozhkov1995}
Stozhkov, Yu. I., Zullo, J. Jr., Martin, I. M., Pellegrino, G. Q.,  et al., 1995. Rainfalls during great Forbush decreases. Il Nuovo Cimento C, 18, 335-341.

\bibitem[\protect\citeauthoryear{Svensmark et al. (2009)}{Svensmark et al., 2009}]{Svensmark2009}
Svensmark, H., Bondo, T., Svensmark, J., 2009. Cosmic ray decreases affect atmospheric aerosols and clouds. Geophys. Res. Lett. 36, L15101.

\bibitem[\protect\citeauthoryear{Tinsley (2008)}{Tinsley, 2008}]{Tinsley2008}
Tinsley, B.A., 2008. The global atmospheric electric circuit and its effects on cloud microphysics. Rep. Prog. Phys. 71, 066801.

\bibitem[\protect\citeauthoryear{Tiwari et al. (2005)}{Tiwari et al., 2005}]{Tiwari2005}
Tiwari, M., Ramesh, R., Somayajulu, B. L. K., Jull, A. J. T.,  Burr, G. S., 2005. Solar control of southwest monsoon on centennial timescales. Curr. Sci. 89, 1583-1588.

\bibitem[\protect\citeauthoryear{Usoskin (2011)}{Usoskin, 2011}]{Usoskin2011}
Usoskin, I. G., 2011. Cosmic rays and climate forcing. Memorie della Societa Astronomica Italiana, 82, 937-942. 

\bibitem[\protect\citeauthoryear{Venkatesan and Badruddin (1990)}{Venkatesan and Badruddin, 1990}]{Venkatesan1990}
Venkatesan, D., Badruddin, 1990. Cosmic ray intensity variations in 3-Dimensional Heliosphere. Space Science Review, 52, 121-194.

\bibitem[\protect\citeauthoryear{Verschuren et al. (2000)}{Verschuren et al., 2000}]{Verschuren2000}
Verschuren, D., Laird, K. R., Cumming, B. F., 2000. Rainfall and drought in equatorial east Africa during the past 1,100 years. Nature, 403, 410-414.

\bibitem[\protect\citeauthoryear{Yadava and Ramesh (2007)}{Yadava and Ramesh, 2007}]{Yadava2007}
Yadava, M. G., Ramesh, R., 2007. Significant longer-term periodicities in the proxy record of the Indian monsoon rainfall. New Astron. 12, 544-555.




\end{thebibliography}
\end{document}